\documentclass{kluwer}

\def\solm{M$_{\odot}\,$}

\usepackage[]{graphicx}
\begin{document}
\begin{article}
\begin{opening}
\title{Observing Massive Galaxy Formation} 
           

\author{\surname{Christopher J. Conselice}\email{cc@astro.caltech.edu}}
\institute{California Institute of Technology, Pasadena CA}         
                      




\runningtitle{Massive Galaxy Formation}
\runningauthor{C.J. Conselice}



\begin{abstract} 
A major goal of contemporary astrophysics is understanding the
origin of the most massive galaxies in the universe, particularly nearby
ellipticals and spirals.  Theoretical models of
galaxy formation have existed for many decades, although 
low and high redshift observations are only beginning to put constraints 
on different ideas.  We briefly describe these
observations and how they are revealing the methods by which galaxies 
form by contrasting
and comparing fiducial rapid collapse and hierarchical formation model
predictions.  The available data show that cluster ellipticals must have
rapidly formed at $z > 2$, and that up to
50\% of all massive galaxies at $z \sim 2.5$ are involved in major mergers.
While the former is consistent with the monolithic collapse picture, we
argue that hierarchal formation is the only model that can reproduce 
all the available observations. 
\end{abstract}




\end{opening}
\vspace{-0.8cm}
\section{Introduction}

Massive galaxies, typically those with stellar masses $> 10^{10}$ M$_{\odot}$,
are the best studied galaxies at all distances due to their high 
luminosities.  Because we can study these galaxies in detail, nearly all 
galaxy formation ideas, models, and scenarios are geared towards constraining
the properties of these systems (Tinsley \& Gunn 1976; Cole et al. 2000). A 
persistent unanswered question is however: how
did these galaxies form?   There are currently no lack of theoretical answers,
although observations are only just beginning to constrain different
ideas.

It is very popular to divide major galaxy formation
scenarios into two different classes: the so-called monolithic collapse
(Tinsley \& Gunn 1976) and hierarchical formation, usually in the milieu of 
the Cold Dark Matter (CDM) paradigm (Cole et al. 2000).   If we 
accept that galaxies are embedded in dark matter halos then  we have an 
additional, and perhaps more fundamental, problem of determining how dark 
halos assembled, and how galaxy formation occurred in concert with this. 

Monolithic collapse models assume that early in the 
universe's history ($z > 2$) baryonic gas in galaxies collapsed to form stars 
within a very
short period of time ($\sim 100$ Myr) with star formation
rates of 10$^{2}$ - 10$^{3}$ \solm yr$^{-1}$,  creating massive galaxies that 
thereafter passively evolve in luminosity.  In hierarchical models, galaxies 
form through mergers of pre-existing smaller systems. 
With these two major ideas as guidelines, we will describe observational
properties of the most massive galaxies in the nearby and distant universe and
how these observations currently favor the hierarchical idea.

Currently we believe that giant elliptical galaxies in clusters formed
their stellar mass early in the universe at $z > 2$ (e.g., Ellis
et al. 1997) and most galaxy
formation studies also try to answer when these objects formed.  
This review is focused on answering the other fundamental origin question of
{\it how} this formation occured.  Three basic pieces of evidence are 
discussed: 1. The fact that nearby giant ellipticals contain old ($> 8$ Gyr) 
stellar populations,
2. The well-defined scaling relations for ellipticals (i.e., fundamental
plane) seen out to $z \sim 1$, and 3.
The structural appearances of high-$z$ galaxies.  While monolithic 
collapse models can account for the first two through a rapid early 
formation, the spatial structures of high-$z$ galaxies at $z \sim 2.5$ can
only be reconciled with the hierarchical model. 
\vspace*{-0.5cm}
\section{Properties of Nearby Massive Galaxies}

Massive galaxies in the local universe include ellipticals,
and spiral galaxies with large central bulges.   
Most galaxy formation models are designed to reproduce the properties of
these systems, which is a natural approach towards understanding 
galaxy formation as nearby systems are the only ones which we can study in
great detail  (see 
Conselice et al. 2002 for a recent overview of low-mass galaxy formation 
which must be dealt with in a different manner.)  We examine
below the properties of nearby giant ellipticals and why most giant
ellipticals in clusters must have formed early.

Ellipticals contain mostly  old stellar populations; this is particularly true 
for those in dense areas (Trager et al. 2000).   The existence 
of these old stellar populations indicate that the bulk of the stars in 
massive galaxies must have formed
long ago.  This is also true to a lesser extent for the bulges of
early type spiral galaxies and ellipticals found in regions of lower
density (Trager et al. 2000).    It appears likely that the oldest stars in 
the universe are in cluster elliptical galaxies.  

To determine the origin of these galaxies, it would be very 
useful to know whether or not the stellar populations in ellipticals formed 
all at once (within a 100 Myr or so), as in the monolithic collapse model, or 
if the stars were formed in discrete episodes. 
Unfortunately, determining the ages of individual stars in these galaxies,
which is currently impossible, 
would not help us resolve this issue, as current techniques for dating old 
$> 5$ Gyr stellar populations are uncertain by at least several Gyrs.  

\begin{figure}
\vspace{-5cm}
\centerline{\includegraphics[width=28pc]{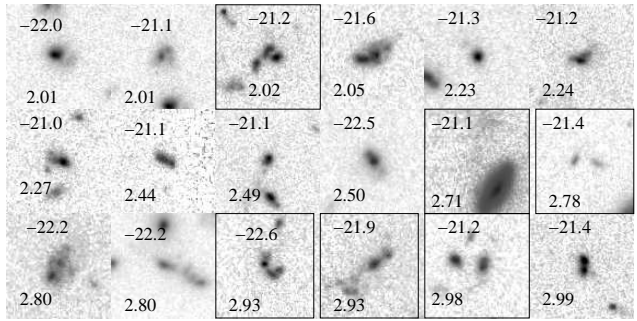}}
\vspace{-4.8cm}
\caption{The brightest, M$_{\rm B} < -21$, galaxies between 2 $< z <$ 3 as 
seen in the Hubble Deep Field North.  Those with asymmetry values
consistent with merging are boxed.  The upper number in each panel is
the M$_{\rm B}$ of each galaxy and the bottom number is its redshift.}
\end{figure}

Other important clues for understanding the origin of elliptical galaxies are
the very strong correlations between photometric and spectroscopic
parameters.  These include the color-magnitude relationship, the correlation 
of velocity dispersion and luminosity, the correlation 
of velocity dispersion and the strength of metal line absorption features, and
the correlation between [$\alpha$/Fe] ratios and velocity dispersion.  These 
correlations, and their tightness out to $z \sim 1$, are all
evidence that giant elliptical galaxies formed very early in the universe
and at nearly the same time, within a few Gyrs (Ellis et al. 1997). The
constraints on exactly when and for how long formation took place are however 
not well known (e.g., Trager et al. 2000).   The only strong 
constraint is that most star formation in cluster ellipticals must have 
occured at $z > 2$. 
\vspace*{-0.5cm}
\section{Properties of High Redshift Galaxies}

\subsection{The Galaxy Population at $z > 2$}

The census of massive high redshift galaxies is likely still incomplete,
but a suitable fraction of the massive high-$z$ population has likely
been uncovered (e.g., Giavalisco 2002).  These include Lyman break 
galaxies (LBGs), extremely red objects (EROs), sub-mm SCUBA sources, and 
quasars.  These systems are likely the progenitors of
the most massive galaxies in the nearby universe, based
on their clustering properties and stellar masses
(e.g,. Papovich et al. 2001; Giavalisco 2002).

\begin{figure}
\centerline{\includegraphics[width=23pc]{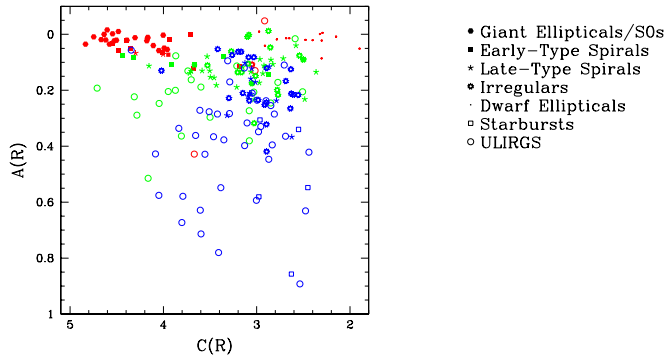}}
\caption{The asymmetry-concentration diagram for nearly all types of
nearby galaxies.  The only systems with high asymmetries are ULRIGs
and starburst undergoing major mergers (from Conselice 2003).  The color
of the point denotes the clumpiness of the galaxy image which correlates
with star formation such that lower to higher values are 
colored red,green,blue.}
\end{figure}

The best characterized of these are the Lyman break galaxies which are high 
redshift $z \sim 2 - 4$ starburst galaxies.  EROs are currently 
thought to be a combination of dusty starbursts or systems
with old stellar populations, with total masses 
$> 10^{12}$ M$_{\odot}$ (Moustakas \& Somerville 2002). SCUBA sources
are likely dusty starbursts, analogs of nearby ULIRGs, that also potentially 
evolve into the large galaxies we see in the nearby universe. 
Not surprisingly, many of these galaxies are
undergoing, or recently underwent, intense star formation, essentially
forming galaxies.  Studying these high-$z$ populations is an extremely active 
research area, although we still only understand their basic 
properties such as current star formation
rates and stellar masses (Papovich et al. 2001).  These
measurements also only reveal when, as opposed to how, these galaxies formed.

\subsection{Structures of High Redshift Galaxies}

A major, and hereto largely unexplored, technique for solving the origin
problem is through the use of structural features, including the
sizes, shapes and morphologies of high redshift galaxies.   One difference 
between young and modern massive galaxies, besides the intensive amounts of 
star formation at high-$z$, is the rarely commented on 
morphological properties of $z \sim 2 - 3$ galaxies as imaged by the Hubble 
Space Telescope (HST) (Figure 1).  These are galaxies that would be 
considered peculiars in most morphology systems, and often mergers from 
the presence of multiple components and tidal tails.  One must
however be careful about making this connection as the appearances
of these galaxies is dominated by intensive star formation at all
wavelengths.  However, recent advances in image analysis  
(e.g., Conselice et al. 2000a;  Conselice 2003) now allow for 
objective morphological methods of identifying galaxies undergoing mergers.

\begin{figure}[H]
\tabcapfont
\centerline{%
\begin{tabular}{c@{\hspace{6pc}}c}
\hspace{2.5cm}
\includegraphics[width=5in]{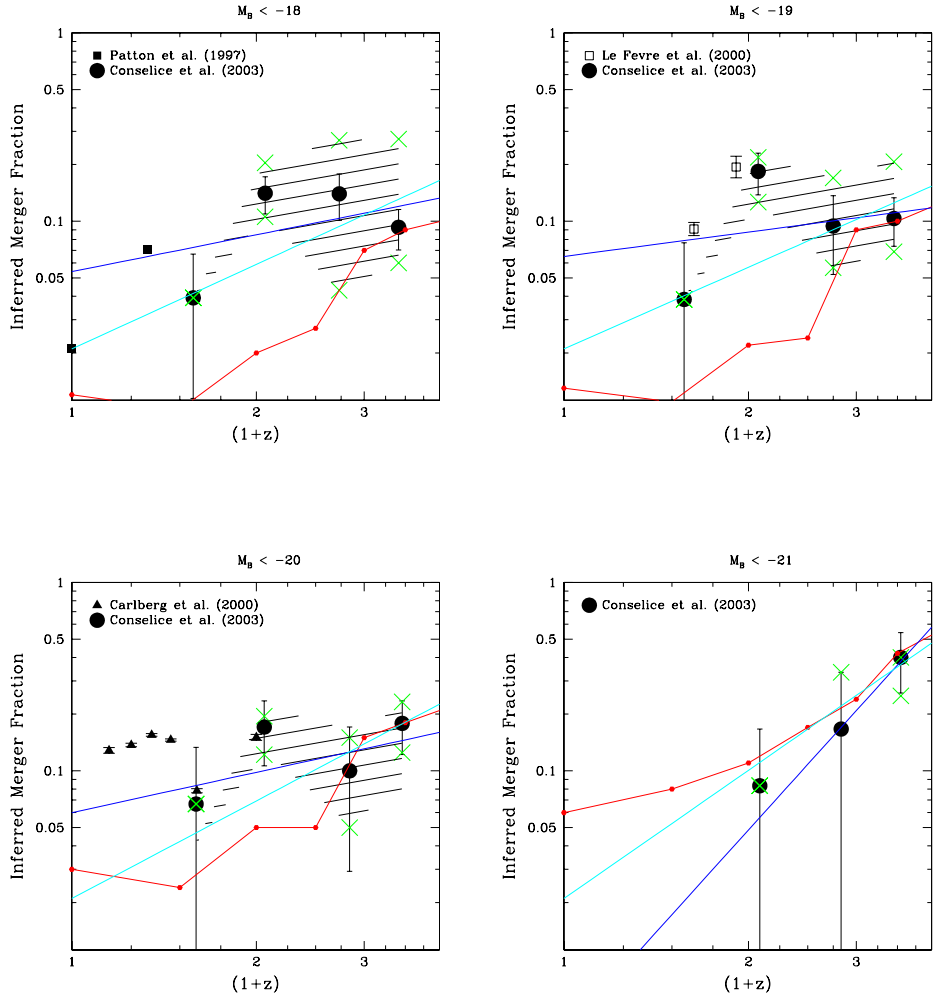} \\
\end{tabular}}
\vspace{0cm}
\caption{The inferred merger fractions of HDF galaxies derived through
the asymmetry method (Conselice et al. 2003).  The solid large circles are 
merger fractions
at different magnitude cuts (shown above each panel) as a function
of redshift.  The other points on the plots are the corresponding
merger fractions derived by Patton et al. (1997), LeF{\'e}vre et al.
(2000) and Carlberg et al. (2000).  The dark blue lines show fits
to only the asymmetry points, while the light blue lines shows similar
fits while holding the $z \sim 0$ fraction constant at 0.02.  The red
line shows merger fraction predictions chosen in exactly the
same way they are observationally, from a semi-analytic CDM model (Benson
et al. 2002). The shaded region is the  $\pm 1\sigma$ random error on the 
merger fractions computations. }
\end{figure}

Most nearby bright galaxies ($> 95$\%) are 
relaxed and can be placed on the Hubble sequence, appearing as simple normal 
ellipticals and spirals.  There are however some nearby galaxies that appear 
peculiar, which cannot be placed into standard classification systems.  A 
subset of these are galaxies undergoing mergers, such as the Ultra
Luminous Infrared Galaxies (ULIRGs).  While the 
process of morphologically classifying galaxies as mergers is
difficult, and subjective, the process is removed of ambiguity by
using a computerized approach (Conselice 2003; Figure~2).

The stellar light distribution of galaxies holds significant information
about the formation histories of galaxies, including whether or not
they recently underwent a major merger (Conselice et al. 2000b; 
Conselice 2003; Figure~2).  Galaxy mergers can be identified 
through the use of the asymmetry index (Conselice et al. 2000a), where the 
most asymmetric galaxies are those that have undergone a major merger in the 
last Gyr (Conselice 2003).   The criteria for choosing a galaxy as a merger
has been found empirically through studies of nearby mergers such
as ULIRGs (Figure~2) and through N-body models 
(Conselice et al. 2003).  Few galaxies, including
those seen at different viewing angles, that are not involved in a recent 
major merger have asymmetry values $> 0.35$.  The only nearby galaxies
with high asymmetries are starbursts and ULIRGs involved in  major 
mergers (Figure~2).  

It is still unknown whether
the application of this criteria at  high redshifts is valid.  Based on
the understanding of how asymmetry behaves for nearby galaxies, it is
a criteria with as much basis as any other, including kinematic 
measurements, as all are based on a priori assumptions of what is a merger.

\begin{figure}[H]
\tabcapfont
\centerline{%
\begin{tabular}{c@{\hspace{5pc}}c}
\hspace{2cm}
\includegraphics[width=4.5in]{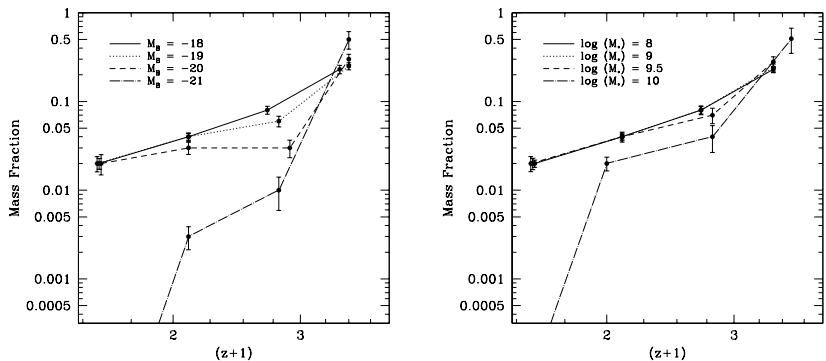} \\
\end{tabular}}
\caption{The fraction of stellar mass involved in major mergers as a function 
of redshift, selected by stellar mass and absolute magnitude limits.  The
most massive M$_{\rm stellar} > 10^{10}$ \solm and brightest systems
M$_{\rm B} < -21$ have the highest fraction (50\%) of their mass involved in 
mergers at $z \sim 2.5$, although this fraction decreases steeply with
redshift.}
\end{figure}

What do we find when we measure the asymmetries of high redshift galaxies in
their rest-frame B-band using WFPC2 and NICMOS observations of the Hubble
Deep Field North? As we would expect in the hierarchical model, there are more
galaxies consistent with mergers at high redshift than at lower redshifts
(Figure~3).  We find that the fraction of galaxies
undergoing major mergers significantly increases for brighter and more
massive systems (Conselice et al. 2003). The fraction of stellar mass 
involved in major mergers also increases steeply with redshift for the most 
massive
and brightest systems (Figure~4). For bright and massive LBGs at $z > 2.5$ 
with M$_{\rm B} < -21$ and M$_{\rm stellar} > 10^{10}$ \solm, approximately
half of all galaxies are involved in major mergers with an evolution of 
$(1+z)^{3.7\pm0.3}$ (Conselice et al. 2003). 

Lower mass and fainter galaxies are not undergoing 
mergers at the same high rate (Figure~3) with an almost constant merger 
fraction history that
only slightly increases at higher redshifts.  The peak merger fraction for
lower mass systems is $\sim$ 20\% at $z \sim 1$.  Our analysis indicates that 
these merger fraction computations are not biased by selection effects or 
systematic errors (Conselice et al. 2003).
\vspace*{-0.5cm}
\section{Testing Models}
\vspace*{-0.2cm}
\subsection{Evidence for Monolithic Collapse}

There are several observations used to argue that massive galaxies
formed through a monolithic collapse, or rather that they did not
form through hierarchical merging.  The main argument in 
favor of an early collapse has to do with what we will call ``smart'' stars 
that inhabit massive galaxies.  Elliptical galaxy stars
are smart because they seem to know the global properties of the
galaxy they inhabit, namely its total luminosity and velocity dispersion 
(\S 2) which likely correlate with total mass.  
The strong correlation between various parameters like [$\alpha$/Fe]
ratios and velocity dispersion, such that more massive systems are
$\alpha$ element enhanced, imply that the giant ellipticals formed
very rapidly, removing gas from these systems before 
significant amounts of Fe from Type Ia supernovae can pollute material
that future generations of stars form from.  Rapid monolithic collapse
models nicely reproduce these features (e.g., Chiosi \& Carrao 2002)

Additionally, if all galaxies form as single systems of the same low mass and 
then merge, the stars inside these  galaxies should all be roughly
homogeneous, if simple
feedback scenarios are implicated for understanding star formation (e.g.,
Dekel \& Silk 1986), and no further star formation occurs and environmental 
effects are ignored.   This is obviously an over idealized situation, but the 
seemingly
uniformly old stellar populations of massive galaxies in clusters at low
and high redshifts (e.g.,
Bower et al. 1992; Ellis et al. 1997) demands an explanation, especially 
since hierarchical
models predict that ellipticals should still be forming up until and later
than $z \sim 1$ (Kauffmann, Charlot \& White 1996).  However, the morphologies
of high-$z$ galaxies (Figure~1 and 3) and the fact that they are not 
homogeneous in
color, but have bluer (and presumably younger) cores (Moth \& Elston 2002;
Menanteau et al. 2001) are strong arguments against a single
monolithic collapse.

\subsection{Evidence for Hierarchical Formation}

As discussed in the introduction, the formation of galaxies through
hierarchical formation is complicated by dark halo formation which may or may 
not be occurring at the same time. To state this another way, dark halos could 
be merging early in the universe, as demanded by CDM models, without any 
cooling of baryons to form stars until most halo merging is finished.  If this
is the case then galaxies form in effectively the same way in monolithic and
hierarchical models.  The question to answer is then whether or not 
massive galaxies formed through mergers of extant galaxies, that is halos 
with stars.

\subsubsection{Clues at Low Redshift}

Observationally, the hierarchical formation of giant galaxies has been
argued based on their steep r$^{1/4}$ surface brightness profiles which
are seen in ongoing mergers such as Arp 220, and are predicted in models
to be the natural outcome of major mergers (Barnes \& Hernquist 1992).   
Perhaps more convincing is the visible
signs of past merging activity around giant elliptical galaxies, such
as the so-called shells or ripples found around 10\% of all massive
galaxies (e.g., Schweizer \& Seizer 1988).    A significant
number of central cluster galaxies also show evidence for recent
merger activity in the form of multiple nuclei and tidal features
(e.g., Conselice et al. 2001).   Another piece of evidence for
merger activity are decoupled cores found in the centers of up to
half of all ellipticals (e.g., de Zeeuw et al. 2002). 
This evidence is however not yet strong enough
to show convincingly that all massive spheroidal galaxies form by 
merging, as not every nearby elliptical has these properties.  Most tell-tale 
structural signatures of merging would also be erased by now, particularly
in dense areas such as clusters. 

\subsubsection{Clues at High Redshift}

One method of determining how massive galaxies formed is to measure the 
mass or luminosity function of galaxies as a function of redshift.  There 
should be more low-mass galaxies and
fewer massive galaxies at high redshift, if mergers occurred, than we see in 
the nearby universe.  At later times the number of massive galaxies
should increase, while the number of lower mass galaxies should decline 
assuming no new galaxy/star formation. 

Another more direct approach is to identify galaxies at high
redshift which are undergoing mergers and to use this to determine the 
past history of merging.  Unfortunately, it is very difficult to
demonstrate with certainly that a given galaxy is undergoing a merger.
One can use identifiable
aspects of nearby major mergers, such as massive star formation, to 
identify mergers, but massive starbursts
can be triggered by a variety of methods (e.g., Conselice
et al. 2000b).   

As we argued in \S 3.2, at least 50\% of galaxies at $z > 2.5$ are 
morphologically consistent with undergoing a merger.  
Other methods of finding mergers through galaxy pair counts, either
kinematic or spatially projected (Patton et al. 1997; LeF{\'e}vre et al. 2000;
Carlberg et al. 2000), agree with our results out to $z \sim 1$ (Figure~3).
As briefly mentioned, there are stellar population gradients 
inside Lyman break galaxies seen in the Hubble Deep Field North such
that the centers of LBGs are bluer than their outer parts (Moth \& Elston 
2002).  This cannot result from a monolithic collapse of gas, and suggests
an origin from preexisting galaxies in merger driven central 
starbursts (Mihos \& Hernquist 1996).

Massive starbursts induced by mergers also solves the smart star problem.
When two dark matter halos merge with galaxies in them,
those galaxies will appear as distinct objects until they merge
due to dynamical friction, although they both occupy effectively the
same dark halo.  Star formation can be induced by the interaction/merger
between these two systems if they contain cold gas.  New stars forming within 
this massive dark halo 
will have properties and characteristics that mimic those of massive galaxies,
producing the required smart stars.   At present we do not directly know if 
the most 
massive nearby galaxies in clusters underwent multiple early episodes of star 
formation, although field ellipticals at $z \sim 1$ show a clear 
diversity of recent star formation properties (Menanteau et al. 2001).
In any case, if a significant fraction of star formation occurs in a massive
galaxy forming through a merger, the resulting star formation will produce
smart stars.

\section{Final Comments}

The implications of these results, with the caveat that the Hubble Deep Field 
North is a small area, is that most massive galaxies formed through the
mergers of preexisting galaxies.   If anything, we are likely underestimating
the fraction of galaxies undergoing mergers at $z \sim 3$ with
the asymmetry method, as fainter and nosier images give systematically lower 
asymmetry values 
(Conselice et al. 2000a).  A similar analysis of the Hubble Deep Field South 
reveals a similar large merger fraction at high redshifts, despite the claim 
that there exist more evolved galaxies at $z \sim 2.5$ (e.g., Labb{\'e} et al. 
2002). Note that although hierarchical formation does seem to be occurring, 
the agreement with semi-analytical CDM models is relatively poor, expect for 
the brightest and most massive systems (Figure~3).  

In the future, deep observations with SIRTF
will help constrain the properties and stellar masses of the underlying older 
stellar populations in these high redshift star forming system.  Kinematic
measurements using IFUs on 8 meter telescopes will also soon reveal
the kinematic properties of these forming galaxies. Ongoing and future 
observations with the Advanced Camera for Surveys on HST will also
allow for a more thorough search and characterization of the merger
process through galaxy structures.

I thank Richard Ellis, Kevin Bundy, Matt Bershady,
Mark Dickinson, Casey Papovich and Mike Santos for valuable 
conversations 
on these topics, and permission to discuss and cite unpublished results.

\end{article}

\vspace{-0.5cm}
\begin{thebibliography}{}
\bibitem[\protect\citeauthoryear{}{}]{} Barnes, J.E., \& Hernquist, L. 1992, ARA\&A, 30, 705
\bibitem[\protect\citeauthoryear{}{}]{} Benson, A. et al. 2002, MNRAS, 333, 156
\bibitem[\protect\citeauthoryear{}{}]{} Bower, R.G., Lucey, J.R., \& Ellis, R.S. 1992, MNRAS, 254, 601
\bibitem[\protect\citeauthoryear{}{}]{} Carlberg, R.G. et al. 2000, ApJ, 532, 1L
\bibitem[\protect\citeauthoryear{}{}]{} Chiosi, C., \& Carrao, G. 2002, MNRAS, 335, 335
\bibitem[\protect\citeauthoryear{}{}]{} Cole, S., Lacey, C.G., Baugh, C.M., \& Frenk, C.S. 2000, MNRAS, 319, 168
\bibitem[\protect\citeauthoryear{}{}]{} Conselice, C.J., Bershady, M.A., \& Jangren, A. 2000a, ApJ, 529, 886
\bibitem[\protect\citeauthoryear{}{}]{}Conselice, C.J., Bershady, M.A., \& Gallagher, J.S. 2000b, A\&A, 354, 21L
\bibitem[\protect\citeauthoryear{}{}]{}Conselice, C.J., Gallagher, J.S., \& Wyse, R.F.G. 2001, AJ, 122, 2281
\bibitem[\protect\citeauthoryear{}{}]{}Conselice, C.J., Gallagher, J.S., \& Wyse, R.F.G. 2002, astro-ph/0210080
\bibitem[\protect\citeauthoryear{}{}]{}Conselice, C.J. 2003, ApJ, submitted 
\bibitem[\protect\citeauthoryear{}{}]{} Conselice, C.J., Bershady, M., Dickinson, M., \& Papovich, C. 2003, in preperation
\bibitem[\protect\citeauthoryear{}{}]{} Dekel, A., \& Silk, J. 1986, ApJ, 303, 39
\bibitem[\protect\citeauthoryear{}{}]{}de Zeeuw, P.T. et al. 2002, MNRAS, 329, 513
\bibitem[\protect\citeauthoryear{}{}]{}Ellis, R. et al. 1997, ApJ, 483, 582
\bibitem[\protect\citeauthoryear{}{}]{}Giavalisco, M. 2002, ARA\&A, 40, 579
\bibitem[\protect\citeauthoryear{}{}]{}Kauffmann, G., Charlot, S., \& White, S.D.M. 1996
\bibitem[\protect\citeauthoryear{}{}]{}Labb{\'e}, I. et al. 2002, astro-ph/0212236
\bibitem[\protect\citeauthoryear{}{}]{} LeF{\'e}vre, O. et al. 2000, MNRAS, 311, 565L
\bibitem[\protect\citeauthoryear{}{}]{} Menanteau, F., Abraham, R.G., \& Ellis, R.S. 2001, MNRAS, 322, 1
\bibitem[\protect\citeauthoryear{}{}]{} Mihos, J.C., \& Hernquist, L. 1996, 464, 641
\bibitem[\protect\citeauthoryear{}{}]{} Moustakas, L.A., \& Somerville, R.S. 2002, ApJ, 577, 1
\bibitem[\protect\citeauthoryear{}{}]{} Moth, P., \& Elston, R.J. 2002, AJ, 124, 1886
\bibitem[\protect\citeauthoryear{}{}]{}Papovich, C., Dickinson, M., \& Ferguson, H.C. 2001, ApJ, 559, 620
\bibitem[\protect\citeauthoryear{}{}]{}Patton, D.R. et al. 1997, ApJ, 475, 29
\bibitem[\protect\citeauthoryear{}{}]{}Schweizer, F., \& Seitzer, P. 1988, ApJ, 328, 88
\bibitem[\protect\citeauthoryear{}{}]{}Tinsley, B.M., \& Gunn, J.E. 1976, 203, 52
\bibitem[\protect\citeauthoryear{}{}]{} Trager, S.C., Faber, S.M., Worthey, G., \& Gonzalez, J.J. 2000, ApJ, 119, 1645

\end{thebibliography}
\end{document}